\documentclass[onecolumn,11pt,floatfix,altaffilletter,superscriptaddress,notitlepage,
     tightenlines,showpacs,showkeys,nofootinbib]{revtex4-1}
\usepackage[utf8]{inputenc}
\usepackage[colorlinks=true,citecolor=blue,linkcolor=blue]{hyperref}
\usepackage[normalem]{ulem}
\usepackage{amsmath,amssymb}
\usepackage{graphicx}               
\usepackage{url}
\usepackage{slashed}
\usepackage{multirow}
\usepackage{enumitem}
\usepackage[dvipsnames]{xcolor}
\usepackage{epstopdf}
\usepackage[capitalise, english]{cleveref}
\usepackage[big,sc,noindentafter]{titlesec}

\makeatletter
\renewcommand{\p@subsection}{}
\renewcommand{\p@subsubsection}{}
\makeatother

\titlespacing*{\section}{0ex}{3ex}{0.5ex}
\titlespacing*{\subsection}{0ex}{2ex}{0.5ex}
\titleformat*{\section}{\normalfont\Large\scshape\color{NavyBlue}}
\titleformat*{\subsection}{\normalfont\large\scshape\color{NavyBlue}}

\begin{document}

\title{Future Opportunities in Accelerator-based Neutrino Physics}

\author{Andrea Dell'Acqua}       \affiliation{CERN, Geneva, Switzerland}
\author{Antoni Aduszkiewicz}     \affiliation{University of Warsaw, Poland}
\author{Markus Ahlers}           \affiliation{Niels Bohr Institute, Copenhagen, Denmark}
\author{Hiroaki Aihara}          \affiliation{University of Tokyo, Japan}
\author{Tyler Alion}             \affiliation{University of Sussex, UK}
\author{Saul Alonso Monsalve}    \affiliation{CERN, Geneva, Switzerland}
\author{Luis Alvarez Ruso}       \affiliation{IFIC (CSIC, University of Valencia), Valencia, Spain}
\author{Vito Antonelli}          \affiliation{University and INFN Milano-Bicocca, Italy}
\author{Marta Babicz}            \affiliation{CERN, Geneva, Switzerland}
                                 \affiliation{Polish Academy of Sciences, Warsaw, Poland}
\author{Anastasia Maria Barbano} \affiliation{University of Geneva, Switzerland}
% ...

\date{December 2018}

\begin{abstract}
This document summarizes the conclusions of the Neutrino Town Meeting 
held at CERN in October 2018 to review the neutrino field at large with the aim of defining a strategy for accelerator-based neutrino physics in Europe.
The importance of the field across its many complementary components is stressed. Recommendations are presented regarding the accelerator based neutrino physics, pertinent to the European Strategy for Particle Physics. We address in particular i) the role of CERN and its  neutrino platform, ii) the importance of ancillary neutrino cross-section  experiments, and iii) the capability of fixed target experiments as well as present and future high energy colliders to search for the possible manifestations of neutrino mass generation mechanisms.  
\end{abstract}


\makeatletter
  \@author@finish
  \title@column%
  \begingroup
      \ltx@footnote@pop
      \def\@mpfn{mpfootnote}%
      \def\thempfn{\thempfootnote}%
      \c@mpfootnote\z@
      \let\@makefnmark\frontmatter@makefnmark
      \frontmatter@setup
      \thispagestyle{titlepage}\label{FirstPage}%
      \frontmatter@title@produce
      \begin{center}
        The Participants of the European Neutrino Town Meeting \\
        22--24 October, 2018 \\
        {\it CERN, 1 Esplanade des Particules, 1211 Geneva 23, Switzerland} \\[0.2cm]
        {\it Editors:}
        Alain Blondel\footnote{alain.blondel@cern.ch},
        Albert De Roeck\footnote{albert.de.roeck@cern.ch},
        Joachim Kopp\footnote{jkopp@cern.ch} \\

        \it (full author list in the appendix)
        \vspace{0.2cm}
      \end{center}
      \frontmatter@RRAPformat{%
       \expandafter\produce@RRAP\expandafter{\@date}%
       \expandafter\produce@RRAP\expandafter{\@received}%
       \expandafter\produce@RRAP\expandafter{\@revised}%
       \expandafter\produce@RRAP\expandafter{\@accepted}%
       \expandafter\produce@RRAP\expandafter{\@published}%
      }%
      \frontmatter@abstract@produce
      \@ifx@empty\@pacs{}{%
       \@pacs@produce\@pacs
      }%
      \@ifx@empty\@keywords{}{%
       \@keywords@produce\@keywords
      }%
      \par
      \frontmatter@finalspace
  \endgroup
  \suppressfloats[t]%
\titlepage@sw{%
  \vfil
  \clearpage
}{}%
\makeatother

\newpage
\section{Foreword}

In order to prepare a contribution to the 2019--2020 Update of the  European Strategy for Particle Physics (ESPP20), based on inputs by the community, and given the particular mission that it received in the 2013 edition, the CERN Neutrino Platform initiated a three-day town meeting, overviewing neutrino physics at large, but aimed at defining a strategy for accelerator-based neutrino physics in Europe. Four panels were created in advance in order to prepare input and conclusions on the key issues. The workshop aims, participation and contributions, as well as the panel membership, missions and reports can be found on the meeting page~\cite{ENTM2018}. 
Discussions at a dedicated  round table and following the panel reports took place. The participants and, the panels, and the round table members should be congratulated for the quality of the scientific discussion. 
  
\section{Recommendations}

\begin{enumerate}[label=\Alph*.,topsep=0pt,itemsep=0pt,parsep=0pt,
                  leftmargin=*,align=left]
\item Neutrino physics is one of the most promising areas where to find answers to some of the big questions of modern physics;
it covers many disciplines of physics complementing each other, and some coordination should ensure that each of these essential aspects is strongly supported.

\item Neutrinos at accelerators, pertinent to ESPP, are an important component because of:
\begin{enumerate}[label=\arabic*),topsep=0pt,itemsep=0pt,parsep=0pt,
                  leftmargin=*,align=left]
  \item the search for CP violation, and  the full
        determination of the oscillation parameters;
  \item the possibility to discover heavy neutrinos or other manifestations          of the mechanism for neutrino mass generation.
\end{enumerate}
Consequently Europe (and CERN in particular) should provide a balanced support in the world-wide LBL effort, with its two complementary experiments DUNE and T2K/Hyper\-Kamiokande (``HyperK'') (and its possible extension with a detector in Korea), in both of which strong EU communities are involved, to secure the determination of oscillation parameters, aim at the discovery of CP violation and test the validity of the 3-family oscillation framework; these experiments also have an outstanding and complementary non-accelerator physics program.

\item Extracting the most physics out of DUNE and HyperK will require ancillary experiments:
\begin{enumerate}[label=\arabic*),topsep=0pt,itemsep=0pt,parsep=0pt,
                  leftmargin=*,align=left]
  \item CERN should continue improving NA61/SHINE towards percent level flux determinations;
  \item a study should be set-up to evaluate the possible implementation, performance and impact of a percent-level electron and muon neutrino cross-section measurement facility (based on e.g.\ ENUBET or NuSTORM) with conclusion in a few years;
\end{enumerate}

\item If, for instance, the CP phase $\delta_{CP}$ is close to $\pm \pi/2$ or of $\sin\delta_{CP}=0 $, improved precision w.r.t.\ DUNE and HyperK should be considered. Studies of feasibility and performance of ESSnuSB and Protvino to Orca (P2O) should be pursued to quantify their feasibility, realistic potential and complementarity with the present program.

\item Fixed target and collider experiments have significant discovery potential for heavy neutrinos and the other manifestations of the neutrino mass generation mechanisms, especially in Z and W decays. The capability to probe massive neutrino mechanisms for generating the matter–-antimatter
asymmetry in  the  Universe  should  be  a  central  consideration  in  the  selection and design of future colliders.

\item A strong theory effort should accompany the experimental endeavours. A specific program to improve Standard Model predictions is needed.
\end{enumerate}

\section{Introduction}

The Physics of massive neutrinos attracts considerable interest by its profound potential implications on the primordial universe and its evolution, as well as its wide range of experimental methods. The bi-annual 2018  Neutrino conference gathered over 800 participants, which is not very different from the 1100 participants of the International Conference on High Energy Physics (ICHEP18). The organization of  neutrino physics, however, span several domains, distributed across different organizational and support frameworks: nuclear physics; astro-particle physics; astronomy; non-accelerator and accelerator physics. The  European Strategy for Particle Physics encompasses a strategy for accelerator-based neutrino physics, traditionally addressed by three type of facilities: neutrino beams, fixed target experiments, and colliders. 

Accelerator-based experiments in Europe have led to important discoveries in the past, from the neutral currents at the CERN PS to the determination of the number of neutrinos at LEP. In the 90's the focus shifted to the search, then the study, of neutrino oscillations. Since 2012, CERN does not host neutrino beams anymore, and the community participates in the long baseline oscillation experiments in the US and in Japan. The highlight objective is to complete the knowledge of oscillation parameters, especially to observe and study CP violation in the neutrino sector. This requires appearance experiments with knowledge of the neutrinos' CP parity -- requirements that can only be met by accelerator-based experiments. European physicists are involved in significant numbers in T2K, its beam intensity upgrade, and in the HyperK project~\cite{Abe:2018uyc}; more recently a large number of Europeans have joined the DUNE experiment in the US~\cite{Abi:2018dnh, Abi:2018rgm, Abi:2018alz}.

DUNE official start was announced in summer 2017, and the University of Tokyo announced recently the start of construction of HyperK in 2020. Both experiments aim to beam data taking with the far detector around 2027.  
These experiments have very different techniques and experimental conditions, and the combination of the two will allow significant cross-checks of both possible sources of errors and of the oscillation paradigm, as well as a most welcome improvement in statistics. DUNE and HyperK offer far deeper complementarity than e.g. the LEP or LHC general purpose detectors. They will also be subject to systematic uncertainties that require ancillary experiments: hadron production experiments for characterizing the neutrino beams, precision studies of neutrino interactions and cross-sections, and energy reconstruction measurements. To all these efforts, which will benefit both DUNE and HyperK, CERN can make essential contributions. 

Furthermore, large underground detectors (40~kton for DUNE and 190~kton for HyperK) offer great and complementary capabilities for proton decay searches, atmospheric neutrinos and supernova detection. The technological developments carried out at the neutrino platform for near or far detectors are synergetic with other fields, such as Dark Matter search experiments for the developments involving giant liquid argon cryostats \cite{Fiorillo}.  Further projects beyond DUNE and HyperK (HyperK--Korea, ESSnuSB, P2O) are under study,  see section~\ref{second-maximum} .  

A unique attractive feature of neutrinos is their electric neutrality, which makes them the only known fermions that admit a Majorana mass term, i.e.\ a mass generation mechanism that transforms neutrinos into antineutrinos. Together with CP violation, this allows them to play a crucial role in the generation of the fermion--antifermion asymmetry of the Universe. Observation of neutrinoless double beta decay would prove fermion number violation, and Europe is at the forefront of this research~\cite{Saakyan, panel2}. Neutrinoless double beta decay should receive highest priority, and a strategy should be developed for probing this process even in the case of normal neutrino mass ordering.  Other observable manifestations of Majorana mass terms can arise, such as the presence of ``sterile'' or ``right-handed'' neutrinos, which might be directly produced in experiments like SHiP and the FCC~(ee, hh,eh). The success of these searches would constitute a transformative breakthrough in the understanding of the mechanism of neutrino masses, opening the possibility to observe in the laboratory the mechanism by which the fermion number asymmetry of the universe is generated; this capacity should be given important consideration in choosing the next high-energy frontier facility, in the same way as the search for dark matter is a flagship goal of the LHC program.

\clearpage
\section{Three-Flavor Oscillation at Long Baseline}
\label{sec:3flavor}

\subsection{Status as of 2018}

Neutrino oscillations demonstrate that neutrinos have mass. This is the most direct laboratory evidence for the existence of physics beyond the Standard Model (SM).
The experimental program aims at measuring the solar, atmospheric and reactor neutrino mixing angles $\sin^2\theta_{12}$,  $\sin^2\theta_{23}$, and $\sin^2\theta_{13}$ with $1\sigma$ relative uncertainties not exceeding approximately 0.7\%, 3\%, and 3\%, respectively, and the CP phase $\delta_{CP}$ with $1\sigma$ uncertainty of approximately $10^\circ$ at $\delta_{CP}\sim 270^\circ$ ~\cite{Petcov:2018}.

For $\sin^2\theta_{13}$, this has already been achieved by the reactor experiments Daya Bay~\cite{An:2012eh}, Reno~\cite{Ahn:2012nd}, and Double Chooz~\cite{Abe:2011fz}. No improvement on this accuracy is expected in the foreseeable future. The current precision on $\sin^2\theta_{12}$ is $\sim 4\%$, and that on $\sin^2\theta_{23}$ is $\sim 9\%$ \cite{Esteban:2016qun}. The error on $\theta_{23}$, which is close to $\pi/4$, is presently large because it is measured by $\nu_\mu$ disappearance, but will improve rapidly as the statistics of $\nu_\mu\rightarrow\nu_e$ events increases. 

The absolute mass squared differences $|\Delta m_{21}^2| \equiv |m_2^2 - m_1^2|$ and $|\Delta m_{31}^2| \equiv |m_3^2 - m_1^2|$, which control the oscillation lengths, are known to 3\% and 1\% accuracy, respectively, $\Delta m_{21}^2$ is known to be positive thanks to the observation of matter effects on solar neutrinos \cite{Mikheyev:1986gs, Wolfenstein:1977ue}, but the sign of $\Delta m_{31}^2$ (commonly referred to as the ``neutrino mass ordering'') \cite{deSalas:2018bym} is still unknown, although T2K \cite{Wascko:2018} and NO$\nu$A \cite{Sanchez:2018} show a weak ($2\sigma$) preference for normal mass ordering ($\Delta m_{31}^2 > 0$) \cite{Esteban:2016qun}. 

Observing CP violation in the lepton sector -- which requires appearance experiments -- would be of paramount interest as it may be related to the matter-antimatter asymmetry of the Universe. CP violation would manifest itself as a difference in the oscillation probability of $\nu_{\alpha} \to \nu_{\beta}$ between neutrinos and antineutrinos. The oscillation channel available for CP violation searches is electron (anti)neutrino appearance in a muon (anti)neutrino beam: $\nu_{\mu} \to \nu_e$ and $\bar{\nu}_{\mu} \to \bar{\nu}_e$. This  is currently probed by T2K and NO$\nu$A, who are already providing first glimpses of CP violation in the lepton sector. Indeed, T2K shows a preference for $\delta_{CP} \sim -\pi/2$ and excludes the CP conserving values $\delta_{CP} = 0, \pi$ at 95\% CL \cite{Wascko:2018}. NO$\nu$A results are compatible with these hints \cite{Sanchez:2018}, which are also reflected by the most recent global fits \cite{Esteban:2016qun}. 
\subsection{The next 5-10 years}
In the coming years, T2K and NO$\nu$A will collect more data, and if the true $\delta_{CP}$ is close to the current global best fit, the expected results from a combined analysis of both experiments could allow for: the exclusion of CP conservation at $>3\sigma$ significance; determination of the neutrino mass ordering at $4\sigma$ significance; a precise determination of $\theta_{23}$ with an uncertainty of 1.7$^\circ$ (3.8\%) or better; a precise measurement of $\Delta m_{31}^2$  with about 1\% precision.

The fully funded JUNO experiment in China~\cite{Djurcic:2015vqa} will start in 2021. The 20~kton liquid scintillator detector located 53~km from the Nuclear power sources, will measure $\Delta m^{2}_{21}$ and sin$^{2}(\theta_{12})$ with 0.6\% and 0.7\% precision, respectively. It is also sensitive to the mass ordering by observing the shorter wavelength atmospheric oscillation,
and should be able to determine the sign of $\Delta m_{31}^2$ with $3\sigma$ significance after 6~years of operation.  

Atmospheric neutrino data are quite sensitive to matter effects. The SuperK experiment released recently an analysis of 16 years of data~\cite{Ikeda:2018}, which  prefers the normal mass ordering at $ > 2\sigma$ level, bringing the present world average preference over $3\sigma$~\cite{Capozzi:2018ubv, deSalas:2017kay,Esteban:2018azc}. The neutrino telescopes KM3NeT--ORCA and PINGU (IceCube Upgrade/Gen2) can provide valuable information on the neutrino mass ordering using matter effects on 2--12~GeV atmospheric neutrinos. 
The expected sensitivity varies between $2\sigma$ and $6\sigma$ (after 3~years of data taking) depending on the true mass ordering and on the value of $\theta_{23}$. ORCA is now partially funded and started deploying the first few PMT lines at sea; full deployment is foreseen to be completed by 2021. PINGU is not yet financed, and deployment could take place between 2025 and 2031~\cite{Katz:2018}. ORCA, SuperK and PINGU, combined with JUNO, T2K and NOVA are likely to provide a determination of the mass hierarchy with a significance of about 5$\sigma$ by the time DUNE and HyperK come online.

\subsection{Future Neutrino Beams: DUNE and Hyper-Kamiokande}

The next-generation long-baseline experiments DUNE~\cite{Abi:2018dnh, Abi:2018rgm, Abi:2018alz} and HyperK~\cite{Abe:2018uyc} will study $\nu_\mu \to \nu_e$ and $\bar\nu_\mu \to \bar\nu_e$ oscillations to search for CP violation, precisely measure the oscillation parameters, search for new physics, and pursue a comprehensive research program using non-accelerator neutrinos. With the DUNE, HyperK, and also JUNO collaborations becoming comparable in size to collider experiments, neutrino physics has taken center stage in the US and Asia. 

In the following discussion the expected performance for DUNE and HyperK is quite sensitive to the systematic error assumptions, which are different for the two projects: HyperK is based on the 2016 oscillation analysis of T2K, while DUNE assumes that errors will scale to match the statistics. 
For DUNE, a Fermilab-based wide-band neutrino beam peaked at 3~GeV will be detected on-axis in four 10~kton fiducial mass Liquid Argon Time Projection Chambers (LArTPC) located 1\,300~km away.  The beam will provide neutrino interactions primarily around the first oscillation maximum, but with a low-energy tail of interactions near the second maximum.  The long baseline allows DUNE to exploit matter effects to determine the mass ordering with $> 5\sigma$ significance in 7~years of operation. Matter effects dynamically generate a neutrino--antineutrino asymmetry that is larger than the maximal effect due to fundamental CP violation. On the other hand, the large matter effects remove approximate parameter degeneracies affecting the determination of the mass ordering and CP phase.  After 10~years of operation, DUNE will be able to discover CP violation at $5\sigma$ (3$\sigma$) for 54\% (74\%) of possible $\delta_{CP}$ values.  Over the same period, $\delta_{CP}$ can be measured with $7.5^{\circ}$ ($15^{\circ}$) precision for $\delta_{CP} = 0$ ($-\pi/2$).  DUNE can determine the octant of $\theta_{23}$ with $> 3\sigma$ significance for values $< 43.5^{\circ}$ or $> 47.9^{\circ}$, and can measure $\theta_{23}$ to a precision of $0.3^{\circ}$ (= 0.7\%) for $\theta_{23}=42^{\circ}$. The anticipated precision of the measurement of $\Delta m_{32}^2$ is 0.3\%.  Both single phase and dual phase detector technology is being developed in the ProtoDUNE prototype modules in operation at the CERN Neutrino Platform. Far detector site construction for DUNE is expected to begin in 2019, with detector installation starting in 2022 and the first detector starting operation in 2024.  The remaining three detector modules will follow over several years. The beam will start operation in 2026 with 1.2~MW beam power, followed by a future upgrade to 2.4~MW.

HyperK will comprise a 186~kton water Cherenkov detector at a distance of 295~km, observing the narrow-band off-axis J-PARC neutrino beam.  The latter is being upgraded to 1.3~MW beam power, so that the beam neutrino event rate in HyperK will be 16 times larger than T2K currently.  Due to the relatively short baseline, the matter-induced neutrino--antineutrino asymmetry is only $\pm 10\%$, while the maximum asymmetry from fundamental CP violation is $\pm 30\%$. 
Given the mass ordering, HyperK will have $5\sigma$ ($3\sigma$) CP violation sensitivity for 57\% (76\%) of possible $\delta_{CP}$ values after 10~years of operation.  Over the same period, the phase can be measured with  $7.2^{\circ}$ ($23^{\circ}$) precision for $\delta_{CP}=0$ ($\delta_{CP}=\pi/2$). HyperK itself will be able to determine the mass ordering ($4\sigma$ after 10~years) by exploiting atmospheric neutrinos. HyperK can determine the octant of $\theta_{23}$ with $> 3\sigma$ significance for values $< 42.7^{\circ}$ or $> 47.3^{\circ}$, and after 10~years of operation can achieve $1^{\circ}$ ($0.5^{\circ}$) degree precision for the $\theta_{23}$ measurements at $\theta_{23}=45^{\circ}$ ($42^{\circ}$ or $48^{\circ}$). Over the same period, HyperK can achieve 0.6\% precision in the measurement of $\Delta m^2_{32}$.  HyperK utilizes established water Cherenkov detection technology, while pursuing improvements to the photo-detectors. Construction of HyperK will begin in 2020, and the detector will be ready for operation in 2027.  Work on the J-PARC beam power upgrade has started, and a beam power of 750~kW (1.3~MW) will be reached in 2021 (2027).

Although these experiments are located outside Europe, European scientists make up 35\% of the DUNE collaboration and 48\% of the HyperK collaboration.

\subsection{Importance of Controlling Systematic Uncertainties}

The excellent statistics available in DUNE and HyperK (e.g.\ $\simeq 2\%$ statistical error on the CP asymmetry) requires control of systematic uncertainties at a level that has never been done before.  Mitigating the uncertainties in the neutrino flux, cross-sections, and event-by-event neutrino energy reconstruction begins with careful design of the near detector complex. Its goal is to i) measure the unoscillated event rate with target isotopes and angular and energy acceptances as similar as possible to the far detector; and to ii) provide the additional information needed for the evaluation of the far detector acceptance and resolution. Ultimately, however, this will not fully eliminate the interaction model dependence~\cite{Alvarez-Ruso:2017oui}. Ancillary experiments, described in \cref{sec:ancillary}, and the associated theoretical efforts, will be essential for the full exploitation of the considerable investment in the long baseline experiments.

\subsection{Complementarity Between Experiments}

We summarize here the synergies and the unique strengths of future neutrino oscillation experiments. JUNO is the only planned experiment offering precision measurements of the ``solar'' oscillation parameters $\Delta m_{21}^2$ and $\theta_{12}$.  Moreover, the combination of JUNO data with results from the already running T2K and NO$\nu$A experiments and from ORCA could lead to a conclusive determination of the neutrino mass ordering in time for DUNE and HyperK.

Regarding the possible discovery of leptonic CP violation and the precision measurement of the parameters $\delta_{CP}$, $\Delta m_{31}^2$, and $\theta_{23}$, the sensitivities of DUNE and HyperK are quite similar.
Nevertheless, given the pivotal importance of systematic uncertainties in these measurements, the availability of two experiments with orthogonal choices regarding beam design, detector technology and baseline will be essential for reaching authoritative conclusions.

The complementarity between DUNE and HyperK becomes even more evident for tests of new physics scenarios: because their relative sensitivities to the standard 3-family oscillation and to non-standard scenarios, such as e.g. a new type of neutrino matter effects  \cite{Liao:2016orc}, are different, their combination will allows to probe and characterize new physics effects more comprehensively.

Finally, DUNE and HyperK are highly complementary in their physics program beyond accelerator-based neutrinos.  With their very different detector technologies, their strengths
in atmospheric neutrino measurements, nucleon decay searches, supernova neutrino detection, high energy astrophysics complement each other.

\subsection{Beyond DUNE and HyperK, Second Oscillation Maximum}
\label{second-maximum}

Current and future neutrino oscillation experiments are optimized to observe oscillations at the first oscillation maximum, where CP violation in the oscillation probabilities is smaller than other terms independent of $\delta_{CP}$. This makes them sensitive to systematic uncertainties. Observations at the second oscillation maximum reduce this problem, the CP violation being relatively larger~\cite{Coloma:2011pg}.

The HyperK collaboration envisions a staged approach with the realization of a second HyperK detector, which could be located in Korea at a baseline of $\sim 1100$~km~\cite{Abe:2016ero}. This is an efficient and straightforwards use of the T2K neutrino beam and provides both qualitative and quantitative improvement in sensitivity which is significantly better than locating a second detector in Kamioka. 

In Europe, the H2020 funded  ESSnuSB\cite{Baussan:2013zcy} design study considers to  use the extremely intense (5~MW) 2~GeV proton linac at the European Spallation Source (ESS) as a neutrino source, aimed at a water Cherenkov detector situated  at 500~km, covering completely and exclusively the second oscillation maximum. It is estimated that ESS$\nu$SB could discover CP violation at $5\sigma$ for up to 60\% of the allowed values of $\delta_{CP}$. The expected precision on $\delta_{CP}$ near $0^\circ$ and $180^\circ$ is about $6^\circ$. This study should continue to reach conclusions on feasibility, cost and performance. 

A study is carried out within the ORCA collaboration to aim a neutrino beam from the accelerator laboratory in Protvino (Russia) to the ORCA neutrino telescope (P2O). The great interest is to avoid underground excavation costs, allowing potentially a much increased detector mass. More detailed studies would be of great interest, so as to understand quantitatively the possible step-wise implementation and ultimate potential of such a set-up. 

\section{Cross Sections and other Ancillary Measurements}
\label{sec:ancillary}

Much of the physics output of DUNE and HyperK will be extracted from the measurement of the appearance probabilities $P(\nu_\mu \to \nu_e)$ and $P(\bar{\nu_\mu} \to \bar{\nu_e})$ as functions of the (anti)neutrino energy. Irreducible uncertainties will arise from the fact that the $\nu_\mu$, $\bar{\nu}_\mu$ $\nu_e$ and $\bar{\nu}_e$ cross-sections and their energy dependence are different, and so are the energy response functions. Ideally, measuring them would require monochromatic beams of known energy, for each of the four neutrino types.   

Theoretically modelling neutrino cross-sections on nuclear targets in the 0.2~GeV to 3~GeV energy region is a formidable challenge. The interplay between different interaction mechanisms, nuclear correlations, meson exchange currents, and binding and excitation energies is in great need of more detailed investigations an measurements, also at low energy transfer and  forward lepton scattering.  The same is true for cross section differences between $\nu_\mu$ and $\nu_e$. At low energy transfers, predictions for differential cross sections are critically different for different calculations. Progress in  this field requires that new and more refined theoretical models should be embedded in generators and experiment Monte-Carlos and compared with more precise data.

\subsection{Hadroproduction Experiments and Beam Characterization}

Long-baseline oscillation experiments will be equipped with near detectors, the primary mission of which is to ensure the normalization of event rates between the near and far location. A proper near-to-far extrapolation requires a good understanding of the phase space distribution of beam neutrinos as well as their flavour composition. This requires precise knowledge of hadron production in the target. The CERN NA61/SHINE experiment provided the data for particles produced in proton--Carbon interactions and in a replica of the T2K target, leading to a flux normalization at the level of  $\pm 5\%$~\cite{Berns:2018tap} both for muon and electron neutrinos. The  experiment now takes data for the NuMI beam line~\cite{Aduszkiewicz:2642286}. Further improvements are possible, using a target tracker for the replica target, by measuring cross-sections for incoming1--10~GeV hadrons, and with an upgrade of the spectrometer and its data acquisition rate. The continuation of NA61/SHINE and a continuous improvement program towards precision measurements for the DUNE and HyperK beams will be an essential contribution from European groups and from CERN.

In order to take full advantage of the available statistics in the far detectors, the near detectors of DUNE and HyperK should cover at least the same scattering angle acceptance as the far detector, and should measure as much of the events as possible, so as to provide input to the far detector simulations. The ND280 off-axis  detector of T2K is being upgraded to this effect with a fine grained (1~cm) 3D scintillating detector (SFGD) surrounded by TPC trackers in a magnetic field to cover the full range of lepton scattering angle.
CERN is very much involved in this project.  

In the T2K set-up, the beam energy spectrum depends on the off-axis angle of the beam. In view of HyperK, an intermediate Water Cherenkov detector is being proposed in order to cover a variety of off-axis beam angles from 1--4 degrees (NUPRISM/E61). This will allow data taking with various different narrow band beams, and the linear combination of these measurements will allow for a much better characterization of the full beam than is currently possible. The DUNE near detector is under design to provide similar features, with the added difficulty of the beam being a wide-band beam; a combination of a liquid argon detector and of a fine-grained detector embedded in a magnet is now considered. A movable or separate detector able to span a range of off-axis angles is also being discussed.

\subsection{Cross-section Experiments}

An important function of near detector systems will be the measurement of neutrino cross-sections and neutrino energy response functions for all relevant species: $\nu_e, \bar{\nu}_e, \nu_\mu, \bar{\nu}_\mu$. A formidable challenge will arise from the fact that electron neutrinos constitute only 1\% of a conventional beam. Moreover, their production reactions are somewhat more complex than those of muon neutrinos. Electron neutrino cross section measurements at the per cent level are likely to require dedicated facilities. 
 
ENUBET is a proposed narrow band beam based on a static focusing system where $\nu_e$s are produced by the three-body semileptonic decay of kaons, and the $\nu_e$ flux at the source is measured at the 1\% level by monitoring large angle positrons in the decay tunnel.

NuSTORM is a muon storage ring design that relies on several decades of R\&D towards future Neutrino Factories. NuSTORM offers similar and very well known fluxes of $\nu_e$, $\bar\nu_e$, $\nu_\mu$, and $\bar\nu_\mu$. On the order of one million interactions of each flavour and CP parity could be produced, with sub-per cent precision in the flux determination.  The tunability of the beam momentum and the precision of the  beam diagnostics make it a very attractive source. In addition, the muon capture and storage techniques developed by NuSTORM could contribute in a decisive manner to the R\&D towards a Muon Collider or the Neutrino Factory proposals. 

Considerable interest was expressed in a dedicated cross-section facility (ENUBET, NuSTORM, or other) which could improve significantly the final precision achievable with DUNE and HyperK. Conceptual Studies should be supported, with the aim to understand the implementation, detector set-up and the comparative physics impact, leading to a conclusion within a few years.

\section{Beyond the Standard Three-Flavor Framework}
\label{sec:bsm}

\subsection{Light Sterile Neutrinos}

Light sterile neutrinos with masses $\leq 100$~eV could participate in neutrino oscillations. A number of $\gtrsim 3\sigma$ anomalies observed in short-baseline oscillation experiments has been  interpreted in this context. This is the case for an excess of events with electromagnetic showers observed in muon (anti)neutrino beams by LSND~\cite{Aguilar:2001ty} and MiniBooNE~\cite{Aguilar-Arevalo:2018gpe}, and for  the $\bar{\nu}_e$ deficit observed in reactors experiments \cite{Mention:2011rk} or using intense radioactive sources \cite{Acero:2007su, Giunti:2010zu}. All these anomalies could have explanations within the SM, but, fault of having a near detector or other way to verify the oscillatory behaviour, no specific source of error has been clearly identified for any of them. Although the data point towards roughly the same parameter region when interpreted in terms of sterile neutrinos, global fits show that a consistent interpretation of \emph{all} anomalies in this framework is not possible \cite{Collin:2016rao, Gariazzo:2017fdh, Dentler:2017tkw, Dentler:2018sju}. 
More specifically, explaining the short baseline $\nu_\mu \to \nu_e$ oscillations in LSND and MiniBooNE necessitates sterile neutrino mixing with muon neutrinos. This has been searched for in a number of experiments, most recently in MINOS/MINOS+ and IceCube; these experiments strongly disfavour at more than $4\sigma$ confidence level the sterile neutrino interpretation of LSND and MiniBooNE. The simplest sterile neutrino models are also excluded by cosmology~\cite{Aghanim:2018eyx}, but numerous theoretical proposals exist for circumventing these constraints, see e.g.~\cite{Hannestad:2013ana, Dasgupta:2013zpn, Chu:2018gxk, Bezrukov:2017ike,Fardon:2003eh}.

In view of this unsatisfactory situation, a comprehensive global program is under way to study the dependence of the anomalies on distance and energy. The goal is to either conclusively prove an oscillation origin for the observed anomalies or to identify sources of error that can explain them.  
Many  proposed or running experiments are situated close to nuclear reactors. For what concerns accelerators, the Fermilab short-baseline program at the Booster beam \cite{Antonello:2015lea} uses liquid argon detectors, MicroBooNE (currently running), ICARUS-T600 (start of operation 2019), and SBND (start of operation in 2020). 
Spanning a distance from 100~m to 600~km, these experiments will test the 99\% parameter regions favored by MiniBooNE and LSND at better than $3\sigma$, and the (smaller) parameter region favored by global fits at $5\sigma$. Importantly, liquid argon detectors can distinguish single  electrons/positrons (oscillation signal) from photons (background from neutral currents in particular)  thus testing one of the leading SM explanations of the MiniBooNE anomaly. 
The JSNS$^2$ experiment at the J-PARC Spallation Neutron Source will use neutrinos from pion decays at rest to verify LSND, which employed a similar source; it expects results in 2021.

\subsection{Heavy Right-Handed Neutrinos at Fixed Target Experiments}

Sterile (or ``right-handed'') neutrinos are a very common prediction of neutrino mass (seesaw) models, but their masses can lie in a very broad range from eV to more than $10^{10}$~GeV. Masses at the Electroweak scale (0.1-100 GeV) are of particular experimental and theoretical interest.  In the range between 100~MeV and a few GeV, they can be produced in weak decays of K, $D$ and $B$ mesons; the small expected mixing angles with the light neutrinos lead to cm to km long decay lengths. High intensity production coupled with a large decay volume is necessary. A beam dump mode of NA62 at CERN is proposed to start investigating the open space from $D$ decays~\cite{Lanfranchi:2017wzl,Drewes:2018gkc}.

The SHiP collaboration proposes to exploit the high energy SPS beam impinging on a beam dump, with a detector of total length of about 50~m, much of it being a decay volume for long lived neutral particles. The experiment expects to improve current limits on right-handed neutrinos by four orders of magnitude, and will cover a significant  fraction of the interesting 
parameter space as guided by cosmology and astrophysics; it will thus have a major impact on the field. 

Other beam dumps that can be explored for this search are the ones used for creating neutrino beams. For instance, the POT delivered to the DUNE/LBNF target will be considerably larger (albeit at lower energy) than the one foreseen for SHiP, and the near detectors can be used to search for right-handed neutrino decay signatures.  The DUNE and HyperK near detectors are, however, not optimized for such searches, and the corresponding sensitivity still has to be evaluated. Also the recent proposal for a large volume surface detector close to an LHC  experiment, named MATHUSLA, has sensitivity to weakly interacting long-lived neutral particles, and can cover a similar region of parameter space compared to SHiP. Quantitative on par comparisons between these different options are required, but there is clearly an excellent experimental potential for the hunt of low mass right-handed sterile neutrinos in the next 10~years.

\subsection{Opportunities at Colliders}

High luminosity hadron and lepton colliders are copious sources of neutrinos via decays of heavy flavor particles and of $W$ and $Z$ bosons, with $W$ and $Z$ decays covering a larger mass range. The mixing of right-handed neutrinos with the three light neutrino spieces may enable their production and decay at colliders. At the LHC, the $W$ production channel has been studied for heavy neutrino production and has led to new limits in the mass--mixing plane. The BELLE II project will cover up to the B mass, and the HL-LHC will cover a competitive search region in the neutrino mass range from 5~GeV up to the $W$ mass. 

The FCC-ee project is particularly favourable for the search for right handed neutrinos in this mass range, owing to the very high luminosity achievable at the $Z$ pole, leading to the production of several 10$^{12}$ $Z$-bosons. This will allow it to observe heavy neutrino decays down to a squared mixing angle of $10^{-11}$, a region of the mass-mixing plane that can generate a baryon asymmetry of the Universe. Precision measurements of processes involving neutrinos 
($Z$ and Higgs invisible widths, tau and W mass, lifetime and branching ratios), can provide indirect evidence for neutrino mixing with heavy states, down to mixing of about $10^{-5}$, but over a much broader mass range. Other neutrino mass models such as the type II and III seesaws or left-right symmetric models involving heavy mediators 
can be detected via lepton number violating signals~\cite{Maiezza:2015lza,Cai:2017mow}. 
These results might guide the design of detectors of the 100 TeV FCC-hh, which, with several orders of magnitude more $W$s than LHC, allowing tagging of flavour and charge both at production and decay of a heavy neutrino, is sensitive to both lepton flavour and fermion number violation.

\section{Neutrinos and the Universe}
\label{sec:universe}

\subsection{Capacities at Future Neutrino Facilities}

Large-scale neutrino detectors DUNE and HyperK (and JUNO for reactors) are multi-purpose observatories.
While physics using neutrino beams is certainly at the center of their
research program, their true potential only becomes evident when also considering
non-accelerator-based data samples, and the combination of the latter with
the former~\cite{Abi:2018dnh,Abe:2018uyc}. We have already discussed the ability to determine the mass ordering with atmospheric neutrinos. Furthermore, thanks to the broad energy and baseline range covered by atmospheric neutrinos, they will significantly
enhance DUNE's and HyperK's ability to test the three family oscillation paradigm. 

When the next galactic supernova explodes, SuperK, JUNO, DUNE, and HyperK are
expected to record thousands of neutrino interactions within a few seconds.
These data will allow for unprecedented insights into the inner workings
of a supernova explosion and the formation of the heavy elements necessary
for life as we know it. Here, the complementarity is particularly important: while HyperK
will provide a precise, time-resolved measurement of the $\bar\nu_e$
flux, DUNE will do the same for the $\nu_e$ flux \cite{Ankowski:2016lab,GalloRosso:2017mdz}. HyperK is also sensitive to supernovae in nearby galaxies.  Further complementarity exists with neutrino telescopes (IceCube, KM3NeT).

JUNO, DUNE, and HyperK are also sensitive 
to nucleon decay, a hallmark signature of Grand Unified Theories (GUTs).
Once again, complementarity is at work: HyperK dominates the sensitivity 
up to proton lifetime over $10^{35}$ years to $p^+ \to e^+ \pi^0$ (mediated for instance by GUT-scale gauge bosons)
thanks to its larger mass \cite{Lopez:1996gy}. JUNO, DUNE and HyperK  
offers sensitivity to $p \to K^+ \nu$ (mediated for instance by
GUT-scale scalars and the superpartners of the light quarks), up to $10^{34}$ years lifetime;
DUNE and JUNO being able to directly identify kaons.

\subsection{Relevance of Non-Accelerator Probes}

The accelerator-based research opportunities that this report is about
should be viewed also in context of the active non-accelerator-based
program in neutrino physics. Experiments pushing the frontiers of
low-background physics to search for neutrinoless double beta decay
($0\nu2\beta$ decay \cite{Vergados:2016hso}) can discover fermion number violation due to the presence of a neutrino Majorana mass term, a prerequisite for explanations of the baryon asymmetry of the Universe and for the seesaw mechanism, which is the leading explanation for the smallness of neutrino masses. The discovery potential is further enhanced in well-motivated extensions of the minimal scenario, such as left-right symmetric models or models with GeV-scale right-handed neutrinos. Among the various experiments worldwide searching for neutrino-less double-beta decay, European experiments such as GERDA (focusing on germanium), CUORE (tellurium) and NEXT (xenon) are some of the most competitive or promising ones~\cite{APPECreport}. The discovery potential of $0\nu2\beta$ decay searches, and the long term strategy for it~\cite{Saakyan}, depend strongly on the determination of the neutrino mass ordering, in which neutrino beam experiments are playing an important role.

$0\nu2\beta$ decay searches can also provide information on the absolute neutrino mass scale, a quantity that oscillations are insensitive to. The least model-dependent measurement of this quantity can be obtained using direct kinematic methods, an effort that is currently led by the tritium-based KATRIN experiment, with an alternative method using Ho-163 being explored by the ECHO and Holmes collaborations. The target sensitivity of these experiments is $\sim 0.2$~eV. Probing even lower masses is the goal of possible future upgrades of KATRIN and of Project~8, which aims to precisely measure the tritium endpoint spectrum by observing the synchrotron radiation emitted by electrons in a magnetic field. Complementary to these laboratory probes, cosmology offers superior indirect sensitivity to neutrino masses. Under the assumption that the $\Lambda$CDM model of cosmology is correct, next generation CMB observatories will begin to probe masses down to $\sim 0.05$~eV, the minimum allowed value inferred from oscillation data.

The emerging field of neutrino astronomy relies on the results of
accelerator-based experiments, for instance for determining neutrino
cross sections and oscillation parameters, and offers complementary
discovery opportunities, in particular in the search for new physics,
and in significantly advancing our understanding of cosmic particle
accelerators.

\section{Role of CERN and the Neutrino Platform}

In 2014, as a response to the recommendations of the 2013 European Strategy Group Report, the Neutrino Platform was established at CERN. 
The current aim of the Platform is
to make essential contributions to the R\&D phase  of future experiments in the short and medium term, and to give coherence to a diverse European neutrino community. The platform provides the community with a test beam infrastructure. It allows for bringing R\&D to the level of technology demonstrators in view of major construction activities. The platform supports the short and long baseline activities for infrastructure and detectors. It acts on demand, using a MOU framework for its activities. The CERN SPSC is the supervising body for the Neutrino Platform: proposals and expressions of interest are submitted for evaluation to this committee. Given a positive review, a recommendation to the CERN Research Board is made for approval and for further project details.
In conjunction with the CERN Neutrino Platform, CERN reinstated an experimental
neutrino group to participate in the world-wide accelerator based neutrino experiments and ancillary experiments. This is the first CERN group having a mandate to participate in off-site experiments.

The Platform acts on direct requests from the community to support 
way new R\&D proposals (hardware and/or software) and to bring in synergies between the neutrino and the LHC community on specific technologies, such as DAQ, computing, electronics. The Platform supports large new projects with the necessary infrastructure available only in big laboratories. The community
can also benefit from the collaborative experience that have been
developed for the LHC.

An example is the LArTPC (time projection chambers) technology that emerged as the solution to be adopted in both short and long baseline experiments in
the US. Two large (700~ton) LArTPC prototypes for the DUNE far detector have been constructed and completed in 2018, and one of them -- the one based on single phase technology -- was used to record millions of
charged particle interactions on argon in fall 2018. New projects being studied include 
the new detector technologies proposed for near detectors for both T2K and DUNE.

The Neutrino Platform aims in particular at assisting the European community for focusing on common R\&D projects. 
The European neutrino community strongly supported the advent of the Neutrino Platform 
in these last years, and expressed a strong wish to see it 
balancing support for projects that are planned both in the eastern and the western part of the world. 

There was also a strong interest expressed from the broader neutrino community to collaborate with CERN
for non-accelerator neutrino projects. Future experiments, e.g.\ on  
the neutrino mass determination, WIMP dark matter or neutrino-less double beta decay searches will see
both detectors and collaborations increase in size, and will be pushing technologies further
to the limit. For many developments (e.g.\ cryogenics, magnets), CERN's expertise would be a strong asset. 

Strong theory support, in which European groups play a leading role, is essential for the success of the future accelerator-based neutrino program, and should be maintained. In the planning stage, theoretical studies are needed to evaluate the scope and physics potential of different detector and beam options.  During data taking, the theoretical interpretation and combination of experimental results is of vital interest. In neutrino physics, a special role has always been played by global fits \cite{deSalas:2017kay, Capozzi:2018ubv, Esteban:2018azc, Collin:2016rao, Gariazzo:2017fdh, Dentler:2018sju}, which often provide the first hints of new effects.  At all stages, precise theoretical predictions of neutrino interactions, implemented in versatile simulation tools, are indispensable. We therefore recommend a stronger involvement of  CERN in theoretical studies, in the same way as CERN's LHC program is accompanied by one of the most visible theory groups in the world.
\newpage



\bibliographystyle{JHEP}
\bibliography{refs}

\providecommand{\href}[2]{#2}\begingroup\raggedright\begin{thebibliography}{10}

\bibitem{ENTM2018}
European Neutrino Town meeting and ESPP 2019 discussion, 22-24 October 2018,
  CERN, \url{https://indico.cern.ch/event/740296/}.

\bibitem{Abe:2018uyc}
{\bf Hyper-Kamiokande} Collaboration, K.~Abe et~al., {\it {Hyper-Kamiokande
  Design Report}},  \href{http://arxiv.org/abs/1805.04163}{{\tt
  arXiv:1805.04163}}.

\bibitem{Abi:2018dnh}
{\bf DUNE} Collaboration, B.~Abi et~al., {\it {The DUNE Far Detector Interim
  Design Report Volume 1: Physics, Technology and Strategies}},
  \href{http://arxiv.org/abs/1807.10334}{{\tt arXiv:1807.10334}}.

\bibitem{Abi:2018rgm}
{\bf DUNE} Collaboration, B.~Abi et~al., {\it {The DUNE Far Detector Interim
  Design Report, Volume 3: Dual-Phase Module}},
  \href{http://arxiv.org/abs/1807.10340}{{\tt arXiv:1807.10340}}.

\bibitem{Abi:2018alz}
{\bf DUNE} Collaboration, B.~Abi et~al., {\it {The DUNE Far Detector Interim
  Design Report, Volume 2: Single-Phase Module}},
  \href{http://arxiv.org/abs/1807.10327}{{\tt arXiv:1807.10327}}.

\bibitem{Fiorillo}
Giuliana Fiorillo, presentation at the European Neutrino Town meeting and ESPP
  2019 discussion, 22 October 2018, CERN,
  \url{https://indico.cern.ch/event/740296/contributions/3160793/attachments/1738688/2812960/Fiorillo_Nu-ESPP.pdf}.

\bibitem{Saakyan}
R. Saakyan, presentation at the European Neutrino Town meeting and ESPP 2019
  discussion, 22 October 2018, CERN,
  \url{https://indico.cern.ch/event/740296/contributions/3171313/attachments/1738530/2812701/Saakyan_0vbb_CernOct18.pdf}.

\bibitem{panel2}
Panel 2 report, European Neutrino Town meeting and ESPP 2019 discussion, 22-24
  October 2018, CERN, \url{https://indico.cern.ch/event/740296/}.

\bibitem{Petcov:2018}
S.~Percov, {\it Neutrino theory including leptogenesis, plenary presentation
  ichep2018, seoul},  July, 2018.

\bibitem{An:2012eh}
{\bf Daya Bay} Collaboration, F.~P. An et~al., {\it {Observation of
  electron-antineutrino disappearance at Daya Bay}},  {\em Phys. Rev. Lett.}
  {\bf 108} (2012) 171803, [\href{http://arxiv.org/abs/1203.1669}{{\tt
  arXiv:1203.1669}}].

\bibitem{Ahn:2012nd}
{\bf RENO} Collaboration, J.~K. Ahn et~al., {\it {Observation of Reactor
  Electron Antineutrino Disappearance in the RENO Experiment}},  {\em Phys.
  Rev. Lett.} {\bf 108} (2012) 191802,
  [\href{http://arxiv.org/abs/1204.0626}{{\tt arXiv:1204.0626}}].

\bibitem{Abe:2011fz}
{\bf Double Chooz} Collaboration, Y.~Abe et~al., {\it {Indication of Reactor
  $\bar{\nu}_e$ Disappearance in the Double Chooz Experiment}},  {\em Phys.
  Rev. Lett.} {\bf 108} (2012) 131801,
  [\href{http://arxiv.org/abs/1112.6353}{{\tt arXiv:1112.6353}}].

\bibitem{Esteban:2016qun}
I.~Esteban, M.~C. Gonzalez-Garcia, M.~Maltoni, I.~Martinez-Soler, and
  T.~Schwetz, {\it {Updated fit to three neutrino mixing: exploring the
  accelerator-reactor complementarity}},  {\em JHEP} {\bf 01} (2017) 087,
  [\href{http://arxiv.org/abs/1611.01514}{{\tt arXiv:1611.01514}}]. NuFit 3.2
  results \url{http://www.nu-fit.org}.

\bibitem{Mikheyev:1986gs}
S.~P. Mikheyev and A.~{\relax Yu}. Smirnov, {\it {Resonance Amplification of
  Oscillations in Matter and Spectroscopy of Solar Neutrinos}},  {\em Sov. J.
  Nucl. Phys.} {\bf 42} (1985) 913--917. [,305(1986)].

\bibitem{Wolfenstein:1977ue}
L.~Wolfenstein, {\it {Neutrino Oscillations in Matter}},  {\em Phys. Rev.} {\bf
  D17} (1978) 2369--2374. [,294(1977)].

\bibitem{deSalas:2018bym}
P.~F. De~Salas, S.~Gariazzo, O.~Mena, C.~A. Ternes, and M.~Tórtola, {\it
  {Neutrino Mass Ordering from Oscillations and Beyond: 2018 Status and Future
  Prospects}},  {\em Front. Astron. Space Sci.} {\bf 5} (2018) 36,
  [\href{http://arxiv.org/abs/1806.11051}{{\tt arXiv:1806.11051}}].

\bibitem{Wascko:2018}
M.~Wascko, {\it {T2K Status, Results, and Plans}}, . Proceedings of the
  Neutrino 2018 Conference, \url{https://doi.org/10.5281/zenodo.1286752}.

\bibitem{Sanchez:2018}
M.~Sanchez, {\it {NOvA Results and Prospects}}, . Proceedings of the Neutrino
  2018 Conference, \url{https://doi.org/10.5281/zenodo.1286758}.

\bibitem{Djurcic:2015vqa}
{\bf JUNO} Collaboration, Z.~Djurcic et~al., {\it {JUNO Conceptual Design
  Report}},  \href{http://arxiv.org/abs/1508.07166}{{\tt arXiv:1508.07166}}.

\bibitem{Ikeda:2018}
M.~Ikeda, {\it Superkamiokande (solar)},  June, 2018.

\bibitem{Capozzi:2018ubv}
F.~Capozzi, E.~Lisi, A.~Marrone, and A.~Palazzo, {\it {Current unknowns in the
  three neutrino framework}},  {\em Prog. Part. Nucl. Phys.} {\bf 102} (2018)
  48--72, [\href{http://arxiv.org/abs/1804.09678}{{\tt arXiv:1804.09678}}].

\bibitem{deSalas:2017kay}
P.~F. de~Salas, D.~V. Forero, C.~A. Ternes, M.~Tortola, and J.~W.~F. Valle,
  {\it {Status of neutrino oscillations 2018: 3$\sigma$ hint for normal mass
  ordering and improved CP sensitivity}},  {\em Phys. Lett.} {\bf B782} (2018)
  633--640, [\href{http://arxiv.org/abs/1708.01186}{{\tt arXiv:1708.01186}}].

\bibitem{Esteban:2018azc}
I.~Esteban, M.~C. Gonzalez-Garcia, A.~Hernandez-Cabezudo, M.~Maltoni, and
  T.~Schwetz, {\it {Global analysis of three-flavour neutrino oscillations:
  synergies and tensions in the determination of $\theta_{23}$, $\delta_{CP}$,
  and the mass ordering}},  \href{http://arxiv.org/abs/1811.05487}{{\tt
  arXiv:1811.05487}}.

\bibitem{Katz:2018}
U.~Katz, {\it Future neutrino telescopes in water and ice}, . Proceedings of
  the Neutrino 2018 Conference, \url{https://doi.org/10.5281/zenodo.1287686}.

\bibitem{Alvarez-Ruso:2017oui}
L.~Alvarez-Ruso et~al., {\it {NuSTEC White Paper: Status and challenges of
  neutrino–nucleus scattering}},  {\em Prog. Part. Nucl. Phys.} {\bf 100}
  (2018) 1--68, [\href{http://arxiv.org/abs/1706.03621}{{\tt
  arXiv:1706.03621}}].

\bibitem{Liao:2016orc}
J.~Liao, D.~Marfatia, and K.~Whisnant, {\it {Nonstandard neutrino interactions
  at DUNE, T2HK and T2HKK}},  {\em JHEP} {\bf 01} (2017) 071,
  [\href{http://arxiv.org/abs/1612.01443}{{\tt arXiv:1612.01443}}].

\bibitem{Coloma:2011pg}
P.~Coloma and E.~Fernandez-Martinez, {\it {Optimization of neutrino oscillation
  facilities for large $\theta_{13}$}},  {\em JHEP} {\bf 04} (2012) 089,
  [\href{http://arxiv.org/abs/1110.4583}{{\tt arXiv:1110.4583}}].

\bibitem{Abe:2016ero}
{\bf Hyper-Kamiokande} Collaboration, K.~Abe et~al., {\it {Physics potentials
  with the second Hyper-Kamiokande detector in Korea}},  {\em PTEP} {\bf 2018}
  (2018), no.~6 063C01, [\href{http://arxiv.org/abs/1611.06118}{{\tt
  arXiv:1611.06118}}].

\bibitem{Baussan:2013zcy}
{\bf ESSnuSB} Collaboration, E.~Baussan et~al., {\it {A very intense neutrino
  super beam experiment for leptonic CP violation discovery based on the
  European spallation source linac}},  {\em Nucl. Phys.} {\bf B885} (2014)
  127--149, [\href{http://arxiv.org/abs/1309.7022}{{\tt arXiv:1309.7022}}].

\bibitem{Berns:2018tap}
{\bf NA61/SHINE} Collaboration, N.~Abgrall et~al., {\it {Measurements of
  $\pi^{\pm}$, $K^{\pm}$ and proton yields from the surface of the T2K replica
  target for incoming 31 GeV/c protons with the NA61/SHINE spectrometer at the
  CERN SPS}},  \href{http://arxiv.org/abs/1808.04927}{{\tt arXiv:1808.04927}}.

\bibitem{Aduszkiewicz:2642286}
{\bf NA61/SHINE Collaboration} Collaboration, A.~Aduszkiewicz, {\it {Report
  from the NA61/SHINE experiment at the CERN SPS}},  Tech. Rep.
  CERN-SPSC-2018-029. SPSC-SR-239, CERN, Geneva, Oct, 2018.

\bibitem{Aguilar:2001ty}
{\bf LSND} Collaboration, A.~Aguilar-Arevalo et~al., {\it {Evidence for
  neutrino oscillations from the observation of anti-neutrino(electron)
  appearance in a anti-neutrino(muon) beam}},  {\em Phys. Rev.} {\bf D64}
  (2001) 112007, [\href{http://arxiv.org/abs/hep-ex/0104049}{{\tt
  hep-ex/0104049}}].

\bibitem{Aguilar-Arevalo:2018gpe}
{\bf MiniBooNE} Collaboration, A.~A. Aguilar-Arevalo et~al., {\it {Significant
  Excess of ElectronLike Events in the MiniBooNE Short-Baseline Neutrino
  Experiment}},  \href{http://arxiv.org/abs/1805.12028}{{\tt
  arXiv:1805.12028}}.

\bibitem{Mention:2011rk}
G.~Mention, M.~Fechner, T.~Lasserre, T.~A. Mueller, D.~Lhuillier, M.~Cribier,
  and A.~Letourneau, {\it {The Reactor Antineutrino Anomaly}},  {\em Phys.
  Rev.} {\bf D83} (2011) 073006, [\href{http://arxiv.org/abs/1101.2755}{{\tt
  arXiv:1101.2755}}].

\bibitem{Acero:2007su}
M.~A. Acero, C.~Giunti, and M.~Laveder, {\it {Limits on nu(e) and anti-nu(e)
  disappearance from Gallium and reactor experiments}},  {\em Phys. Rev.} {\bf
  D78} (2008) 073009, [\href{http://arxiv.org/abs/0711.4222}{{\tt
  arXiv:0711.4222}}].

\bibitem{Giunti:2010zu}
C.~Giunti and M.~Laveder, {\it {Statistical Significance of the Gallium
  Anomaly}},  {\em Phys. Rev.} {\bf C83} (2011) 065504,
  [\href{http://arxiv.org/abs/1006.3244}{{\tt arXiv:1006.3244}}].

\bibitem{Collin:2016rao}
G.~H. Collin, C.~A. Argüelles, J.~M. Conrad, and M.~H. Shaevitz, {\it {Sterile
  Neutrino Fits to Short Baseline Data}},  {\em Nucl. Phys.} {\bf B908} (2016)
  354--365, [\href{http://arxiv.org/abs/1602.00671}{{\tt arXiv:1602.00671}}].

\bibitem{Gariazzo:2017fdh}
S.~Gariazzo, C.~Giunti, M.~Laveder, and Y.~F. Li, {\it {Updated Global 3+1
  Analysis of Short-BaseLine Neutrino Oscillations}},  {\em JHEP} {\bf 06}
  (2017) 135, [\href{http://arxiv.org/abs/1703.00860}{{\tt arXiv:1703.00860}}].

\bibitem{Dentler:2017tkw}
M.~Dentler, A.~Hern\'{a}ndez-Cabezudo, J.~Kopp, M.~Maltoni, and T.~Schwetz,
  {\it {Sterile neutrinos or flux uncertainties? — Status of the reactor
  anti-neutrino anomaly}},  {\em JHEP} {\bf 11} (2017) 099,
  [\href{http://arxiv.org/abs/1709.04294}{{\tt arXiv:1709.04294}}].

\bibitem{Dentler:2018sju}
M.~Dentler, A.~Hern\'{a}ndez-Cabezudo, J.~Kopp, P.~A.~N. Machado, M.~Maltoni,
  I.~Martinez-Soler, and T.~Schwetz, {\it {Updated Global Analysis of Neutrino
  Oscillations in the Presence of eV-Scale Sterile Neutrinos}},  {\em JHEP}
  {\bf 08} (2018) 010, [\href{http://arxiv.org/abs/1803.10661}{{\tt
  arXiv:1803.10661}}].

\bibitem{Aghanim:2018eyx}
{\bf Planck} Collaboration, N.~Aghanim et~al., {\it {Planck 2018 results. VI.
  Cosmological parameters}},  \href{http://arxiv.org/abs/1807.06209}{{\tt
  arXiv:1807.06209}}.

\bibitem{Hannestad:2013ana}
S.~Hannestad, R.~S. Hansen, and T.~Tram, {\it {How secret interactions can
  reconcile sterile neutrinos with cosmology}},  {\em Phys.Rev.Lett.} {\bf 112}
  (2014) 031802, [\href{http://arxiv.org/abs/1310.5926}{{\tt
  arXiv:1310.5926}}].

\bibitem{Dasgupta:2013zpn}
B.~Dasgupta and J.~Kopp, {\it {A m\'enage \`a trois of eV-scale sterile
  neutrinos, cosmology, and structure formation}},  {\em Phys.Rev.Lett.} {\bf
  112} (2014) 031803, [\href{http://arxiv.org/abs/1310.6337}{{\tt
  arXiv:1310.6337}}].

\bibitem{Chu:2018gxk}
X.~Chu, B.~Dasgupta, M.~Dentler, J.~Kopp, and N.~Saviano, {\it {Sterile
  Neutrinos with Secret Interactions -- Cosmological Discord?}},
  \href{http://arxiv.org/abs/1806.10629}{{\tt arXiv:1806.10629}}.

\bibitem{Bezrukov:2017ike}
F.~Bezrukov, A.~Chudaykin, and D.~Gorbunov, {\it {Hiding an elephant: heavy
  sterile neutrino with large mixing angle does not contradict cosmology}},
  \href{http://arxiv.org/abs/1705.02184}{{\tt arXiv:1705.02184}}.

\bibitem{Fardon:2003eh}
R.~Fardon, A.~E. Nelson, and N.~Weiner, {\it {Dark energy from mass varying
  neutrinos}},  {\em JCAP} {\bf 0410} (2004) 005,
  [\href{http://arxiv.org/abs/astro-ph/0309800}{{\tt astro-ph/0309800}}].

\bibitem{Antonello:2015lea}
{\bf MicroBooNE, LAr1-ND, ICARUS-WA104} Collaboration, M.~Antonello et~al.,
  {\it {A Proposal for a Three Detector Short-Baseline Neutrino Oscillation
  Program in the Fermilab Booster Neutrino Beam}},
  \href{http://arxiv.org/abs/1503.01520}{{\tt arXiv:1503.01520}}.

\bibitem{Lanfranchi:2017wzl}
{\bf NA62} Collaboration, G.~Lanfranchi, {\it {Search for Hidden Sector
  particles at NA62}},  {\em PoS} {\bf EPS-HEP2017} (2017) 301.

\bibitem{Drewes:2018gkc}
M.~Drewes, J.~Hajer, J.~Klaric, and G.~Lanfranchi, {\it {NA62 sensitivity to
  heavy neutral leptons in the low scale seesaw model}},  {\em JHEP} {\bf 07}
  (2018) 105, [\href{http://arxiv.org/abs/1801.04207}{{\tt arXiv:1801.04207}}].

\bibitem{Maiezza:2015lza}
A.~Maiezza, M.~Nemevsek, and F.~Nesti, {\it {Lepton Number Violation in Higgs
  Decay at LHC}},  {\em Phys. Rev. Lett.} {\bf 115} (2015) 081802,
  [\href{http://arxiv.org/abs/1503.06834}{{\tt arXiv:1503.06834}}].

\bibitem{Cai:2017mow}
Y.~Cai, T.~Han, T.~Li, and R.~Ruiz, {\it {Lepton Number Violation: Seesaw
  Models and Their Collider Tests}},  {\em Front.in Phys.} {\bf 6} (2018) 40,
  [\href{http://arxiv.org/abs/1711.02180}{{\tt arXiv:1711.02180}}].

\bibitem{Ankowski:2016lab}
A.~Ankowski et~al., {\it {Supernova Physics at DUNE}},  in {\em {Supernova
  Physics at DUNE Blacksburg, Virginia, USA, March 11-12, 2016}}, 2016.
\newblock \href{http://arxiv.org/abs/1608.07853}{{\tt arXiv:1608.07853}}.

\bibitem{GalloRosso:2017mdz}
A.~Gallo~Rosso, F.~Vissani, and M.~C. Volpe, {\it {What can we learn on
  supernova neutrino spectra with water Cherenkov detectors?}},  {\em JCAP}
  {\bf 1804} (2018), no.~04 040, [\href{http://arxiv.org/abs/1712.05584}{{\tt
  arXiv:1712.05584}}].

\bibitem{Lopez:1996gy}
J.~L. Lopez, {\it {Supersymmetry: From the Fermi scale to the Planck scale}},
  {\em Rept. Prog. Phys.} {\bf 59} (1996) 819--865,
  [\href{http://arxiv.org/abs/hep-ph/9601208}{{\tt hep-ph/9601208}}].

\bibitem{Vergados:2016hso}
J.~D. Vergados, H.~Ejiri, and F.~Šimkovic, {\it {Neutrinoless double beta
  decay and neutrino mass}},  {\em Int. J. Mod. Phys.} {\bf E25} (2016), no.~11
  1630007, [\href{http://arxiv.org/abs/1612.02924}{{\tt arXiv:1612.02924}}].

\bibitem{APPECreport}
European Astroparticle Physics Strategy 2017-2026,
  \url{http://www.appec.org/roadmap}.

\end{thebibliography}\endgroup

\section*{Author List}

\makeatletter
\def\frontmatter@authorformat{\raggedright}

\def\@affil@script#1#2#3#4{%
 \@ifnum{#1=\z@}{}{%
  \par
  \begingroup
   \frontmatter@affiliationfont
   \@ifnum{\c@affil<\affil@cutoff}{}{%
    \def\@thefnmark{#1}\@makefnmark
   }%
   \raggedright
   \ignorespaces\mbox{#3}%
   \@if@empty{#4}{}{\frontmatter@footnote{#4}}%
   \par
  \endgroup
 }%
}%

\begingroup
  \ltx@footnote@pop
  \def\@mpfn{mpfootnote}%
  \def\thempfn{\thempfootnote}%
  \c@mpfootnote\z@
  \let\@makefnmark\frontmatter@makefnmark
  \frontmatter@setup
  \frontmatter@author@produce@script
\endgroup

\makeatother

\end{document}